\documentclass[article,reprint,amsmath,amssymb,raggedbottom,letterpaper,final,superscriptaddress,citeautoscript,floatfix,notitlepage]{revtex4-2}
\usepackage{graphicx}% Include figure files
\usepackage{dcolumn}% Align table columns on decimal point
\usepackage{bm}% bold math

\usepackage[usenames,dvipsnames]{color}
\usepackage{graphicx,microtype}
\usepackage{ulem}

\begin{document}

\preprint{APS/123-QED}

\title{Coupling of Magnetic Phases at Nickelate Interfaces}
 
\author{C. Domínguez}
\email{Corresponding author: claribel.dominguez@unige.ch}
\affiliation{DQMP, University of Geneva, Geneva, Switzerland}
\author{J. Fowlie}
\affiliation{Stanford Institute for Materials and Energy Sciences, SLAC National Accelerator Laboratory, Menlo Park, CA, USA}
\affiliation{Department of Applied Physics, Stanford University, Stanford, CA, USA}
\author{A. B.\ Georgescu}
\affiliation{Department of Materials Science and Engineering, Northwestern University, Evanston, Illinois, USA}
\author{B. Mundet}
\affiliation{DQMP, University of Geneva, Geneva, Switzerland}
\affiliation{LSME, École Polytechnique Fédérale de Lausanne (EPFL), Lausanne, Switzerland}
\author{N. Jaouen}
\affiliation{Synchrotron SOLEIL, Gif-sur-Yvette, France}
\author{M. Viret}
\affiliation{SPEC, CEA, CNRS, Université Paris-Saclay, Gif-sur-Yvette, France}
\author{A. Suter}
\affiliation{Paul Scherrer Institut, Laboratory for Muon Spin Spectroscopy, Villigen PSI, Switzerland}
\author{A. J.\ Millis}
\affiliation{Center for Computational Quantum Physics, Flatiron Institute, New York, USA}
\affiliation{Department of Physics, Columbia University, New York, USA}
\author{Z. Salman}
\affiliation{Paul Scherrer Institut, Laboratory for Muon Spin Spectroscopy, Villigen PSI, Switzerland}
\author{T. Prokscha}
\affiliation{Paul Scherrer Institut, Laboratory for Muon Spin Spectroscopy, Villigen PSI, Switzerland}
\author{M. Gibert}
\affiliation{Solid State Physics Institute, TU Wien, Vienna, Austria}
\author{J.-M. Triscone}
\affiliation{DQMP, University of Geneva, Geneva, Switzerland}

\begin{abstract}
In this work we present a model system built out of artificially layered materials, allowing us to understand the interrelation of magnetic phases with that of the metallic-insulating phase at long length-scales, and enabling new strategies for the design and control of materials in devices. The artificial model system consists of superlattices made of SmNiO$_3$ and NdNiO$_3$ layers – two members of the fascinating rare earth nickelate family, having different metal-to-insulator and magnetic transition temperatures. By combining two complementary techniques- resonant elastic x-ray scattering and muon spin relaxation- we show how the magnetic order evolves, in this complex multicomponent system, as a function of temperature and superlattice periodicity. We demonstrate that the length scale of the coupling between the antiferromagnetic and paramagnetic phases is longer than that of the electronic metal-insulator phase transition - despite being subsidiary to it. This can be explained via a Landau theory- where the bulk magnetic energy plus a gradient cost between magnetic and non magnetic phases are considered. These results provide a clear understanding of the coupling of magnetic transitions in systems sharing identical order parameters. 
\end{abstract}

\maketitle

%\tableofcontents

\section{\label{sec:level1}Introduction}

Rare earth nickelates with the general formula RNiO$_3$ (R = rare earth) are an interesting family of materials that are characterized by a strong interplay between charge, spin, orbital and lattice degrees of freedom. In RNiO$_3$ this results in a metal-insulator transition (MIT) and a complex magnetic order at low temperatures  – except for LaNiO$_3$ that is paramagnetic and metallic at all temperatures \cite{RN223,RN224}. Beyond these well documented properties, multiferroicity has been predicted in the RNiO$_3$ low temperature insulating antiferromagnetic phase \cite{RN330} and superconductivity has recently been observed in infinite layer nickelates upon reduction of the perovskite phase \cite{RN256,RN328}. 

The RNiO$_3$ present a distorted structure from the ideal cubic, with the level of structural distortion of the perovskite crystal depending on the rare earth size, with the NiO$_6$ octahedra tilted and rotated to fill the extra space around the rare earth ion. These rotations cause the pseudocubic unit cell to be smaller, modifying the electronic overlap between the Ni \textit{d} and O \textit{p} orbitals. The temperature of the metal-insulator transition is strongly linked to the Ni-O-Ni bond angle and thus the rare earth. The MIT occurs simultaneously with a lowering of the crystal symmetry from the high temperature orthorhombic metallic phase (\textit{Pbnm} symmetry)  to the low temperature monoclinic insulating phase (\textit{P2$_1$/n} symmetry) \cite{RN27} wherein two inequivalent Ni sites are established (note that LaNiO$_3$, the only member of the family not displaying a MIT, is rhombohedral), with the electron-lattice coupling playing a key role in the transition \cite{RNLandscape}. In the monoclinic phase, one set of NiO$_6$ octahedra is compressed with short Ni-O bonds and the other set of octahedra is expanded with long Ni-O bonds forming a periodic arrangement of alternating large and small NiO$_6$ octahedra referred to as bond disproportionation (BD) \cite{RN243,RN227,RN142,RN113,RN105}. The BD is coupled to a disproportionation of charge (or charge density) - in the extreme case $d_6$ + $d_8$ in the low energy picture or $d^8L^2$ + $d^8$, where \textit{L} designates a ligand hole, when Ni \textit{d} and O \textit{d} hybridization is considered \cite{RN27,RN105,RN365}. In the antiferromagnetic insulating phase, the $d^8L^2$ Ni shares two electrons with the surrounding oxygens that are weakly spin-polarized and the $d^8$ Ni has two electrons in the e$_g$ orbital in a high spin configuration \cite{RN12,RN135}. In the extreme case this leads to a spin ordering of S = 0 and S = 1 respectively. The system orders antiferromagnetically within the insulating phase for T $<$ T\textsubscript{Néel} with no further orbital ordering \cite{RN121}. The antiferromagnetic phase is identified by the pseudocubic (pc) Bragg vector q\textsubscript{Bragg} = $\left( \frac{1}{4},\frac{1}{4},\frac{1}{4} \right)$\textsubscript{pc}, which indicates a periodicity of 4 Ni planes along the [111] direction (or $\left( \frac{1}{2},0,\frac{1}{2} \right)$ in \textit{Pbnm} notation) \cite{RN170,RN100,RN277}. For smaller rare earths (Sm and onwards), T\textsubscript{Néel} is lower than T\textsubscript{MI}  \cite{RN100,RN277}. For large rare earths (Nd and Pr), the system orders magnetically at the same temperature as the bond and charge ordering and T\textsubscript{MI} coincides with T\textsubscript{Néel}. This reflects the quenching of the magnetic order by electronic fluctuations in the metallic state \cite{RN170,RN100,RN277,RN101}.

Most of the recent studies of this family of materials have been performed on heterostructures as single crystals remain available only at the micron scale \cite{RN131,RN216}. The ability to grow heterostructures can also give rise to a plethora of properties that are absent in the bulk counterparts. Emergent magnetic phenomena can, for instance, arise due to interfacial boundary conditions or dimensionality effects. As an example, an induced antiferromagnetic order can be stabilized along the [111] pseudocubic direction in superlattices made of ultrathin LaNiO$_3$ – paramagnetic in bulk form – and LaMnO$_3$ – an antiferromagnet in bulk \cite{RN162}. In a similar architecture but in the [001] direction, superlattices alternating a few unit cells of LaNiO$_3$ and the wide-gap insulator LaAlO$_3$ are found to display, with decreasing temperature, a metal-to-insulator and a paramagnetic-antiferromagnetic transition \cite{RN272}. Also, both collinear and non-collinear magnetic structures can be realized in [111]\textsubscript{pc}-oriented NdNiO$_3$ slabs depending on their thickness \cite{RN202}.

Here we present a study on superlattices made of SmNiO$_3$ and NdNiO$_3$ layers with the thickness of the repeating unit, the superlattice wavelength, denoted $\Lambda$. In bulk, a T\textsubscript{MI} $\sim$ 400 K and $\sim$ 200 K is observed for SmNiO$_3$ and NdNiO$_3$ respectively. When these two compounds are brought together at an interface the stability of a metal-insulator phase separation can be controlled by the thickness of the individual layers, leading to a critical length scale ($\Lambda_{(c-MIT)} = 16$ unit cells (u.c.)) below which- a single metal-to-insulator transition occurs \cite{RN290}. Transmission electron miscroscopy in combination with electron energy-loss spectroscopy (STEM-EELS) confirms that below $\Lambda_{(c-MIT)}$ the entire structure is metallic at room temperature whereas above $\Lambda_{(c-MIT)}$ the individual nickelate layers are sharply separated into metallic (NdNiO$_3$) and insulating (SmNiO$_3$) phases \cite{RN340}. We demonstrated that this behavior is set by the balance between the energy of the interfacial electronic and lattice mismatch at the phase boundary and the bulk phase energies \cite{RN290}.

As the ground state of these compounds is not only insulating but also antiferromagnetic, this leads to questions on the cost of magnetic phase boundaries, specifically the antiferromagnetic-paramagnetic phase boundary and the possible coupling of the magnetic transitions. As the magnetic phase is often treated as subsidiary to the insulating phase, the magnetic critical length scale may be expected to coincide with the electronic one, particularly in NdNiO$_3$ where the two transition temperatures coincide in bulk. On the other hand, antiferromagnetism has recently been suggested to influence the mechanism of the MIT on a local scale and a theory involving the coupling of the magnetic and charge order parameters was introduced to describe this \cite{RN166}. Our current study shows that the length scales of the magnetic transition are different to those of the metal-insulator transition, an effect that is most likely due to the different transition types: while the metal-insulator transition is first-order, the magnetic transition is second order, leading to longer-range behavior and a smoother transition both in temperature and in the length scales of the coupling.

To study magnetism in these artificial structures, we have combined two complementary techniques - resonant elastic x-ray scattering and muon spin relaxation - to show how the magnetic order evolves in SmNiO$_3$/NdNiO$_3$ superlattices as a function of temperature and superlattice periodicity. To understand the experimental results, we expand on the Landau theory from our previous work \cite{RN290}, describing the metal-insulator transition and its length scales in superlattices considering an electronic disproportionation order parameter and a magnetic order parameter coupled to it, described by a second order transition. As a result a complete phase diagram for T\textsubscript{MI} and T\textsubscript{Néel} has been established experimentally and understood theoretically. 

Similar to the metallic-insulating phase boundary, an energy cost between magnetically ordered and non-magnetically ordered phases gives rise to a second bifurcation point where these phases coexist. The critical length scale over which an antiferromagnetic-paramagnetic phase coexistence can occur is found to be greater than the critical length scale for insulating-metallic phase coexistence, and the transition significantly smoother; as a result, non-zero magnetic order may be found in the metallic phase of NdNiO$_3$, as induced by SmNiO$_3$. Our simple model system allows us to better understand the coupling of the metal-to-insulator and magnetic transitions in systems sharing identical order parameters.

\begin{figure*}
    \centering
    \includegraphics [width=\textwidth]{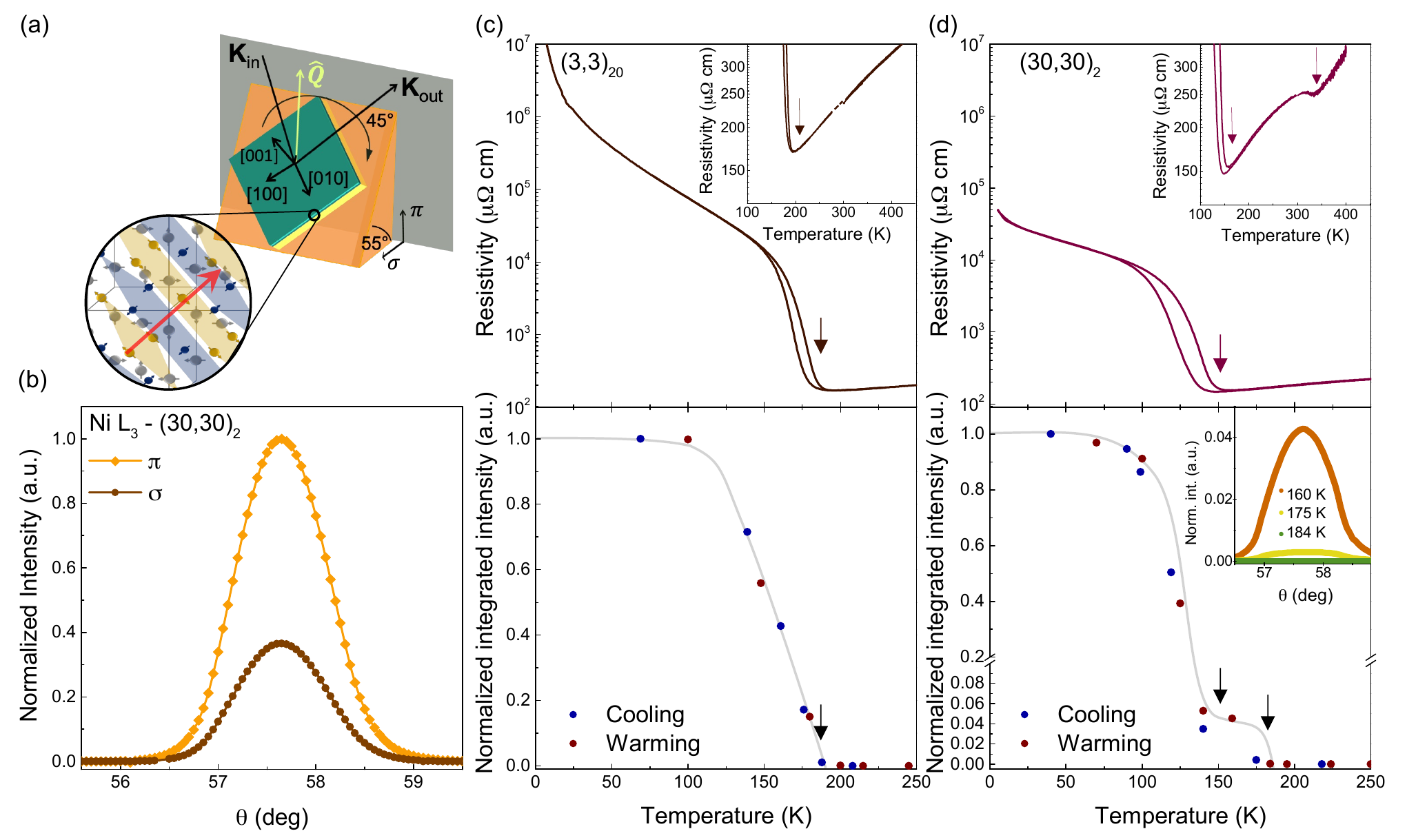}
    \caption{Resonant elastic X-ray scattering (REXS) experiments. (a) Schematic representation of the diffraction geometry used for the magnetic X-ray scattering. Samples were mounted rotated 45° in-plane on a 55° tilted wedge. Also shown are the wave vectors of the incoming and outgoing photons (black arrows) where the orientation of the diffraction plane is colored gray and the incoming $\pi$ and $\sigma$ photon polarizations. The complex antiferromagnetic ordering of the rare earth nickelate family is also shown with the wavevector represented by the red arrow, adapted from reference \cite{RN198}. (b) Rocking curve of the antiferromagnetic $\left( \frac{1}{4},\frac{1}{4},\frac{1}{4} \right)$\textsubscript{pc} Bragg peak measured at 30 K with incoming photon energy in resonance with the Ni-L$_3$ edge (E = 852.5 eV) for $\pi$ (orange) and $\sigma$ (brown) polarizations. Resistivity and intensity of the $\left( \frac{1}{4},\frac{1}{4},\frac{1}{4} \right)$\textsubscript{pc} Bragg reflection as a function of temperature for (c) a short-period (3,3)$_{20}$ and (d) a long-period (30,30)$_2$ superlattice. The inset in the lower part of panel (d) shows rocking curves about q\textsubscript{Bragg} = $\left( \frac{1}{4},\frac{1}{4},\frac{1}{4} \right)$\textsubscript{pc}, at T = 160, 175 and 184 K. Note that the lower part of the intensity scale in panel (d) has been expanded for clarity. The solid gray lines serve as guides to the eye. The arrows indicate transition temperatures. Error bars representing one standard deviation from a Gaussian fit are around the size of the data points.}
    \label{rexsm}
\end{figure*}

\section{METHODS}

Epitaxial ((SmNiO$_3$)$_m$/(NdNiO$_3$)$_m$)$_L$ superlattices, where \textit{m} is in pc u.c. and \textit{L} indicates the number of repetitions of the basic unit ((SmNiO$_3$)$_m$/(NdNiO$_3$)$_m$) were deposited on top of [001]\textsubscript{pc}-oriented LaAlO$_3$ substrates, using radiofrequency off-axis magnetron sputtering as previously described \cite{RN290}. The total heterostructure thickness was kept at $\sim$40 nm, and $\Lambda = 2m$ denotes the superlattice wavelength. All superlattices are shown to have excellent quality and grow coherently strained to the substrate with atomically sharp interfaces and very limited cationic intermixing between the NdNiO$_3$ and SmNiO$_3$ layers \cite{RN290}. To fully determine the magnetic phase diagram of this superlattice system, and to understand how T\textsubscript{Néel} and T\textsubscript{MI} are related, we used resonant elastic x-ray scattering and muon spin relaxation techniques. The former is particularly well-suited for studying magnetic ordering of epitaxial superlattices. The latter can accurately measure the magnetic volume fraction. Thus, by combining a probe of long-range magnetic order and a highly sensitive probe of local magnetism, the difficulty to distinguish antiferromagnetic order as well as to disentangle the effects of loss of coherency with a loss of magnetic volume can be overcome.\\ 
The resistivities of the superlattices were acquired as a function of temperature between 4–400 K. Below 300 K, the samples were slowly dipped into a liquid helium bath, whereas measurements made above room temperature were achieved using two Peltier elements. The measurements were performed in a conventional Van der Pauw geometry with Pt contacts sputtered on the corners of the samples.

\section{Experimental Results}
\subsection{\label{sec:level2}Resonant Elastic X-Ray Scattering}

To probe the existence of the $\left( \frac{1}{4},\frac{1}{4},\frac{1}{4} \right)$\textsubscript{pc} antiferromagnetic order of the SmNiO$_3$/NdNiO$_3$ superlattices, we performed linearly polarized resonant elastic x-ray scattering (REXS) experiments at the RESOXS \cite{RN262,RN352} diffractometer SEXTANTS beamline end-station of the SOLEIL synchrotron in France. Measurements were done at the Ni-L$_{2,3}$ edges and at the Nd-M$_5$ and Sm-M$_5$ edges between 30 and 300 K using a continuous-flow helium cryostat. All REXS measurements were done in absence of magnetic field. To access the $\left( \frac{1}{4},\frac{1}{4},\frac{1}{4} \right)$\textsubscript{pc} reflection of our heterostructures, the (001)\textsubscript{pc}-oriented superlattices were mounted rotated 45$^{\circ}$ in-plane on a 55$^{\circ}$ tilted copper wedge as shown in the schematic representation in Figure \ref{rexsm} (a). At T = 30 K a sharp diffraction peak is observed in a (30,30)$_2$ superlattice, see Figure \ref{rexsm} (b). The peak position corresponds to the 4 monolayer vector oriented along the [111] direction. The diffraction peak is observed in all the superlattices measured and resonates when the incoming beam energy is tuned to the Ni-L$_{2,3}$ edges confirming that the signal has a non-structural origin. An example of the energy scan can be found in Figure S1 in the supplementary material. As the intensity at the L$_3$ edge is larger in magnitude than the intensity at the L$_2$ edge, we analyzed our data when the incoming photon energy is resonant to the Ni-L$_3$ edge. By rotating the polarization of the incoming beam from horizontal ($\pi$) to vertical ($\sigma$) a strong dichroism is observed between the $\pi$ and $\sigma$ channels (Figure \ref{rexsm} (b)). Because of the resonant nature of the peak, the observed dichroic effect and the clear similarity to the bulk and thin film cases, we can state that the detected Bragg reflection is of magnetic origin \cite{RN62,RN260} and that the antiferromagnetically-ordered phase of the rare earth nickelates is also stabilized in the superlattices.

Figures \ref{rexsm} (c) and (d) show the electrical resistivity (top panel) together with the evolution of the diffracted intensity of the $\left( \frac{1}{4},\frac{1}{4},\frac{1}{4} \right)$\textsubscript{pc} magnetic Bragg reflection (bottom panel) characteristic of the antiferromagnetic order as a function of temperature for a (3,3)$_{20}$ short-period and a (30,30)$_2$ large-period superlattice, respectively. At each temperature, the intensity data points were obtained by integrating the rocking curve scans. The blue (red) points correspond to the cooling (warming) cycles. The peak intensity at the Ni-L$_3$ resonance decreases as the temperature increases, thus, the antiferromagnetic-paramagnetic transition temperature (T\textsubscript{Néel}) of our superlattices has been defined as the temperature at which the diffraction intensity peak vanishes. Since we use this definition for T\textsubscript{Néel}, the metal-to-insulator transition temperature is determined from the resistivity measurements by taking the minimum of the resistivity \footnote{Note that the metal-to-insulator transition temperature (T\textsubscript{MI}) can also be defined as the temperature of the maximum in the plot $-d(lnR)/dT$ against T upon heating \cite{RN213}, like in our previous publication \cite{RN290}.}.

\begin{figure}[h!]
\includegraphics[width=\columnwidth]{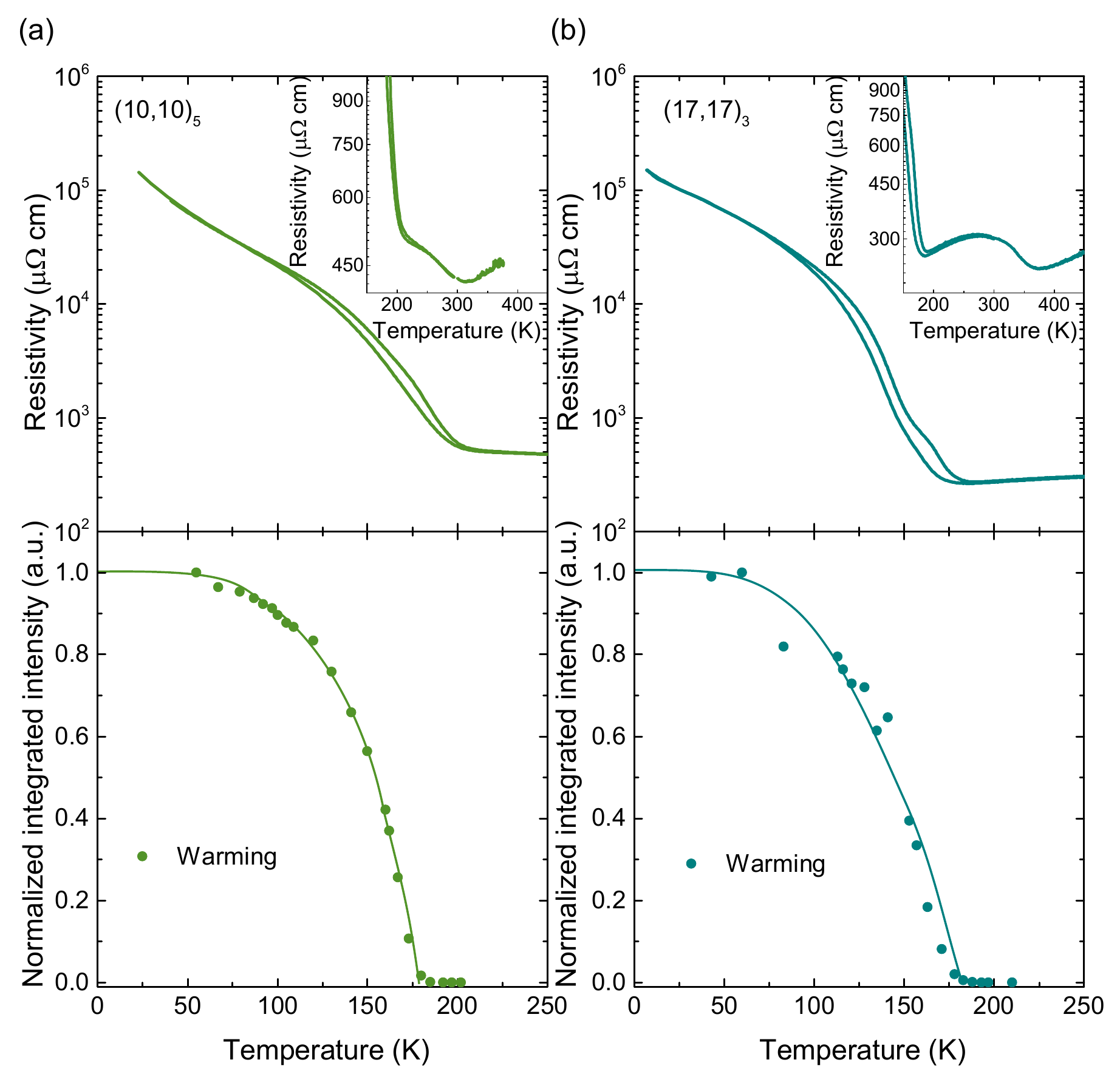}
\caption{Resistivity and intensity of the $\left( \frac{1}{4},\frac{1}{4},\frac{1}{4} \right)$\textsubscript{pc} Bragg reflection as a function of temperature for (a) (10,10)$_{5}$ and (b) (17,17)$_3$ superlattices. The solid lines serve as guides to the eye. The arrows indicate transition temperatures. Error bars representing one standard deviation from a Gaussian fit are around the size of the data points.}
\label{rexsm_extra} 
\end{figure}

As shown in Figure \ref{rexsm} (c) top panel, the short-period (3,3)$_{20}$ superlattice displays a single insulator-to-metal transition with significant hysteresis at $\sim$ 190 K, indicative of a first-order phase transition. From REXS shown in the bottom panel of Figure \ref{rexsm} (c), it is concluded that the onset of the magnetic transition coincides approximately with the unique insulator-to-metal transition observed in this superlattice system, resembling the evolution of the antiferromagnetically-ordered phase observed in bulk and thin film NdNiO$_3$, with T\textsubscript{Néel} = T\textsubscript{MI} \cite{RN121,RN170}. As can be seen in Figure \ref{rexsm} (d) top panel, for the long-period (30,30)$_2$ superlattice two insulator-to-metal transitions occur, a first one at 150 K and a second one at around 350 K, resembling those of individual NdNiO$_3$ and SmNiO$_3$ thin films \cite{RN163,RN217}. Notably, the temperature dependence of the magnetic peak intensity (bottom panel) shows a clear drop accompanied by a pronounced kink at the lowest insulator-to-metal transition (150 K). The peak intensity does not fully disappear at this temperature but continues decreasing until it vanishes completely at around 180 K (inset bottom panel (d)). The short-period behavior has also been observed for a (7,7)$_8$ superlattice and, likewise, the long-period behavior in another long-period superlattice (25,25)$_2$ (not shown here). However, for intermediate superlattices periods such as (10,10)$_5$ and (17,17)$_3$, despite displaying two separate MITs, a single magnetic transition was observed occurring together with the lowest insulator-to-metal transition temperature, see Figure \ref{rexsm_extra}. Specifically with T\textsubscript{Néel} at $\sim$ 190 K for (10,10)$_5$ and T\textsubscript{Néel} at $\sim$ 180 K for (17,17)$_3$, fairly close to the T\textsubscript{Néel} value reported for SmNiO$_3$ films \cite{RN217}. All these results are summarized on Figure \ref{diagram}, to which we will return shortly.

Additionally, we looked at the temperature dependence of the $\left( \frac{1}{4},\frac{1}{4},\frac{1}{4} \right)$\textsubscript{pc} reflection at the Nd-M$_5$ and Sm-M$_5$ edges of a (17,17)$_3$ superlattice. The diffraction peak is observed and starts to resonate at a lower temperature than the Ni L$_{2,3}$, demonstrating that the antiferromagnetism of the Ni sublattice induces the magnetic response of the rare earth sublattice, see Figure S2 (b) and (c), as has been observed previously in bulk \cite{RN100,RN43} and nickelate thin films \cite{RN217,RN309}.

\begin{figure}[h!]
\includegraphics[width=\columnwidth]{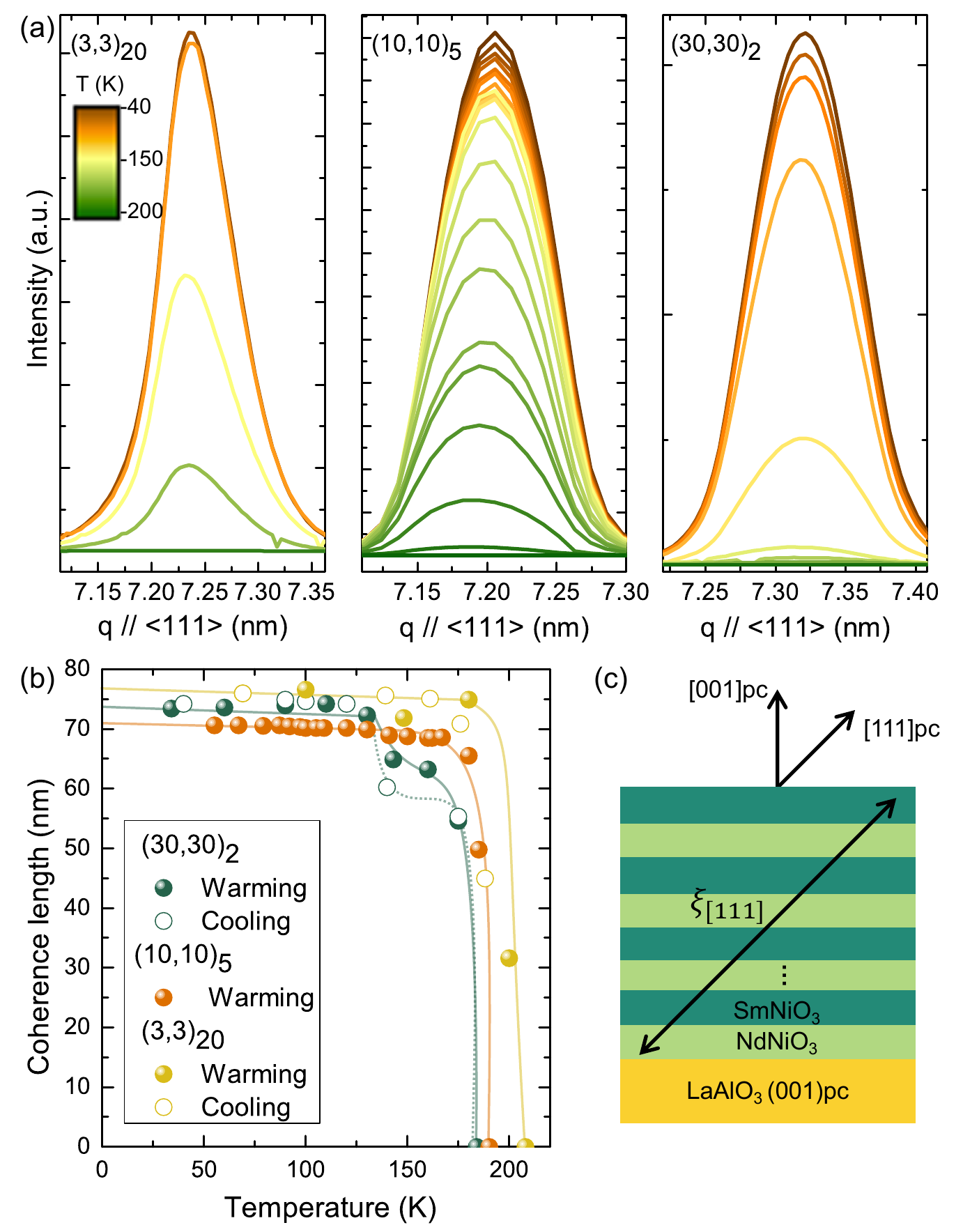}
\caption{Magnetic coherence length. (a) Resonant scattered intensity for three superlattices (3,3)$_{20}$, (10,10)$_5$, and (30,30)$_2$ measured at different temperatures. (b) Scherrer coherence length of (3,3)$_{20}$, (10,10)$_5$ and (30,30)$_2$ superlattices as a function of temperature. (c) Schematic 2-D representation of the superlattices depicting the measured coherence length along the [111] direction. The solid/dotted lines in (b) serve as guides to the eye. Error bars representing one standard deviation from a Gaussian fit are around the size of the data points.}
\label{coherence}
\end{figure}

To gain further insight, we use the Scherrer formula to calculate the coherence length contributing to the magnetic Bragg peak from REXS measurements for three different superlattices: (3,3)$_{20}$, (10,10)$_5$ and (30,30)$_2$ and plot it as a function of temperature, see Figure \ref{coherence} (b). By converting the FWHM (Full width half maximun, in reciprocal lattice units) to a coherence length, $\xi = 2\pi/FWHM$, one can determine the “grain” size – the coherent size of the scattering array probed at the \textit{Q} vector. For (3,3)$_{20}$ and (10,10)$_5$, we find that the coherence length at low temperatures is 76 nm and 70 nm, respectively. These values are comparable to the total film thickness along the [111] direction- 77 nm and 66 nm, respectively- suggesting that the coherent size is limited only by the thickness of the films. The coherence length remains constant for all the set of temperatures measured, indicating that the antiferromagnetic wave vector is coherent through the entire superlattice until T $\sim190$ K for the (3,3)$_{20}$ and (10,10)$_5$ superlattices (Figure \ref{coherence} (b)). This behavior is attributed to the single magnetic transition observed in these heterostructures. For the (3,3)$_{20}$ superlattice it has been proved that the coherence length versus temperature behavior is comparable upon cooling and warming.

For a (30,30)$_2$ superlattice, the value of the coherence length found at low temperatures is 73 nm, also comparable to the total film thickness along the [111] direction, which is 77.5 nm. At $\sim$150 K, the temperature above which the NdNiO$_3$ layers become metallic, the coherence length decreases to 60 nm. This is significantly larger than the total thickness of the two SmNiO$_3$ layers along the [111] direction. The thickness of each SmNiO$_3$ layer in this direction is 19.4 nm, making the total SmNiO$_3$ thickness equal to 38.8 nm, much smaller than the coherence length of 60 nm. This observation suggests that, at these intermediate temperatures, the $\left( \frac{1}{4},\frac{1}{4},\frac{1}{4} \right)$ antiferromagnetic order present in the SmNiO$_3$ layers might still be coherent in most of the superlattice through the neighboring NdNiO$_3$ layers. The observed behavior might be due to a gradual propagation of the magnetic order from one material into the next, as the theoretically predicted magnetic profile of a (30,30)$_2$ superlattice shows in Fig 6. Notice that, the coherence length values are also comparable upon cooling and warming for the (30,30)$_2$ superlattice.

\subsection{Low Energy Muon Spin Relaxation}

Muons have a large magnetic moment, which makes them sensitive magnetic probes. The muon spin relaxation technique ($\mu$SR) exploits this strong magnetic sensitivity to evaluate how the muon spin polarization of an ensemble of muons implanted, one at a time, in a sample evolves with time. This technique is made possible thanks to the unique properties of the muon decay. When the positive muon ($\mu^+$) decays it emits a positron preferentially along the muon spin direction. By measuring the anisotropic distribution of the decay positrons from an ensemble of muons implanted, one can determine the statistical average direction of the spin polarization of the muon ensemble.

The $\mu$SR experiments were carried out using the unique low energy muon (LEM) beam at the $\mu$E4 beamline at the Paul Scherrer Institute in Villigen, Switzerland. Here, the LEM instrument produces low-energy muons with tunable energies in the range of 1 to 30 keV. This allows access to different implantation depths in solids and thin films, ranging from a fraction of a nm to several hundred nm \cite{RN367,RN287Dunsiger,RN282Need,RN349Fowlie}. For our superlattices, simulated muon implantation profiles were calculated for various implantation energies with the TRIM.SP Monte Carlo code \cite{RN366,RN312}. All the experiments were carried out using an implantation energy of 3 keV. The calculated probability of $\mu^+$ shows that for this energy the muons stop in the middle of the heterostructure, an example can be found in Figure S4 in the supplementary material for a (45,45)$_2$ superlattice. The samples were measured over the temperature range 10 - 300 K using a gas-flow cryostat and spectra were recorded both upon cooling and warming.

\begin{figure*}
    \centering
    \includegraphics [width=\textwidth]{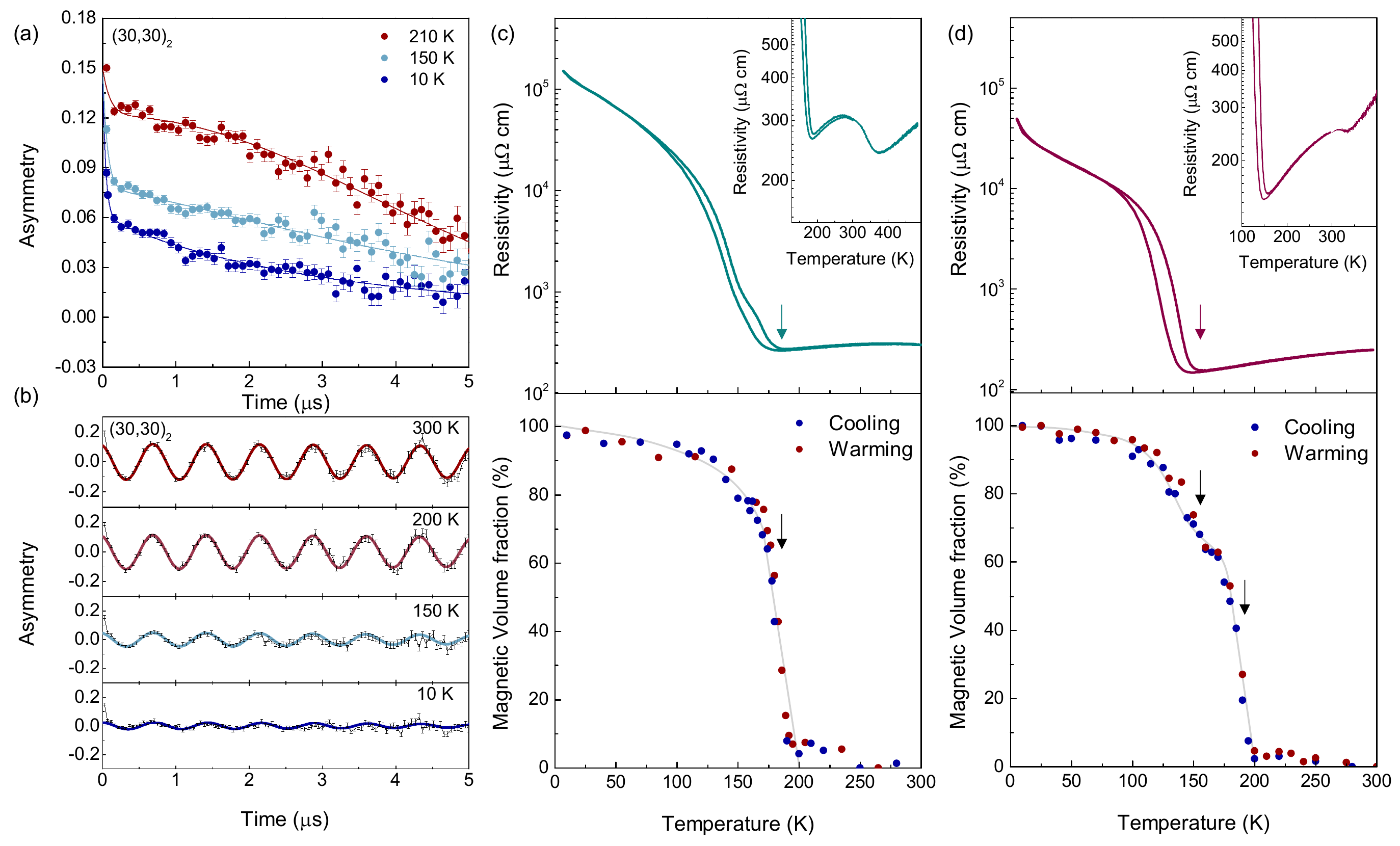}
    \caption{Muon spin relaxation ($\mu$SR) experiments. (a)  Zero-field and (b) Weak transverse magnetic field (wTF) time spectra at various temperatures for a (30,30)$_2$ superlattice. The colored dots represent the data, the solid curves the fits. Resistivity and temperature dependence of the magnetic volume fraction extracted from the wTF data for (c) (17,17)$_3$ and (d) (30,30)$_2$ superlattices. The solid gray lines serve as guides to the eye. The arrows indicate transition temperatures. Error bars representing one standard deviation from the fit are around the size of the data points}
    \label{muons}
\end{figure*}

We performed $\mu$SR measurements in zero externally applied magnetic field (ZF) and in a weak transverse magnetic field (wTF) applied perpendicular to both the initial direction of the muon spin polarization and the film surface. These measurements were carried out for (10,10)$_5$, (17,17)$_3$, (30,30)$_2$ and (45,45)$_2$ superlattices at various temperatures, providing a detailed picture of the evolution of the magnetic ordering temperature in the superlattices as a function of the superlattice period. Figure \ref{muons} (a) shows the time evolution of the ZF muon spin polarization at selected temperatures (10 K, 150 K and 210 K) for a (30,30)$_2$ superlattice. The ZF time spectra exhibit a monotonic decay with no coherent oscillations. However, significant differences are found above and below the lowest metal-to-insulator transition of this sample (T\textsubscript{MI} = 150 K). For the whole temperature range, the zero field asymmetry signal of the positron decay spectra can be described by the following function (solid lines in Figure \ref{muons} (a)):

\begin{equation}
    A(t) = A_Fe^{-\lambda_Ft} + A_S(e^{(-\lambda_St)})^\beta
\end{equation}

\noindent
where $A_F$ and $A_S$ are the asymmetries for the fast and slow depolarizing parts, respectively. For $\beta \cong 2$, and $A_F \cong 0$, the Gaussian-like $A(t)$ with a relatively slow depolarization rate, $\lambda_S$, is typical for a paramagnetic or diamagnetic state that is governed by nuclear dipole broadening, here from the nuclear moments of Sm and Nd.  This is found for temperatures higher than 200 K (where all the layers are paramagnetic, as confirmed by the REXS data). As the temperature decreases, the asymmetry, $A_S$, drops sharply, and $A(t)$ changes to a more bi-exponential behavior, i.e. $\beta \cong 1$, and $A_F \neq 0$. Here, $\lambda_F$ and $\lambda_S$ are measures of the distribution of local static and dynamic fields. This is typical behavior below a magnetic transition. This signature could have two potential causes: (i) the internal field distribution is much larger than the zero field precession signal, or (ii) the zero field precession signal is too high to be observed. For LEM the upper limit of observable ZF precession signals is $\sim$60MHz.

Figure \ref{muons} (b) displays the muon spin wTF time spectra in a field B = 10 mT for a (30,30)$_2$ superlattice. The solid lines represent a decaying cosine fit:

\begin{equation}
A(t) = A_0e^{-\lambda_{TF}t} cos(\gamma_\mu B t + \phi) 
\end{equation}

\noindent
where $\gamma_\mu$ is the gyromagnetic ratio of the muon, $\phi$ a detector phase, and $\lambda_{TF}$ the transverse field depolarization rate. In wTF measurements on magnetic materials, the amplitude $A_0$ is proportional to the non-magnetically ordered volume fraction; thus, a spectrum with negligible oscillation (such as the one at T = 10 K) corresponds to a fully ordered sample, while a spectrum with the full asymmetry (such as the one at T = 300 K) indicates a completely paramagnetic or diamagnetic sample. Note that the very small residual oscillation amplitude at T = 10 K is due to some background contributions.

Figure \ref{muons} (c) and (d) show the resistivity and magnetic volume fraction,

\begin{equation*}
    f_M (T) = 1 - A_0 (T)/A_0 (300 K)
\end{equation*}

\noindent 
expressed as a percentage and extracted from the amplitude of the wTF data as a function of temperature for the (17,17)$_3$ and (30,30)$_2$ superlattices, respectively. As can be seen, both superlattices display two separate insulator-to-metal transitions; see corresponding insets in the top panels. As shown in Figure \ref{muons} (c), the (17,17)$_3$ superlattice shows a sharp reduction in the magnetic volume fraction at the same temperature at which the lowest insulator-to-metal transition is observed. In the case of a (30,30)$_2$ superlattice as the magnetic volume fraction decreases a first kink is observed at the lowest insulator-to-metal transition (160 K), until it vanishes completely at around 190 K. Respectively, a similar type of behavior is observed for a (10,10)$_5$ and (45,45)$_2$ superlattice (see Figure S5 in the supplementary material). These results are consistent with the REXS experiments shown in the previous section and are summarized on Figure \ref{diagram}.

\section{Discussion}

\begin{figure}[h!]
\includegraphics[width=\columnwidth]{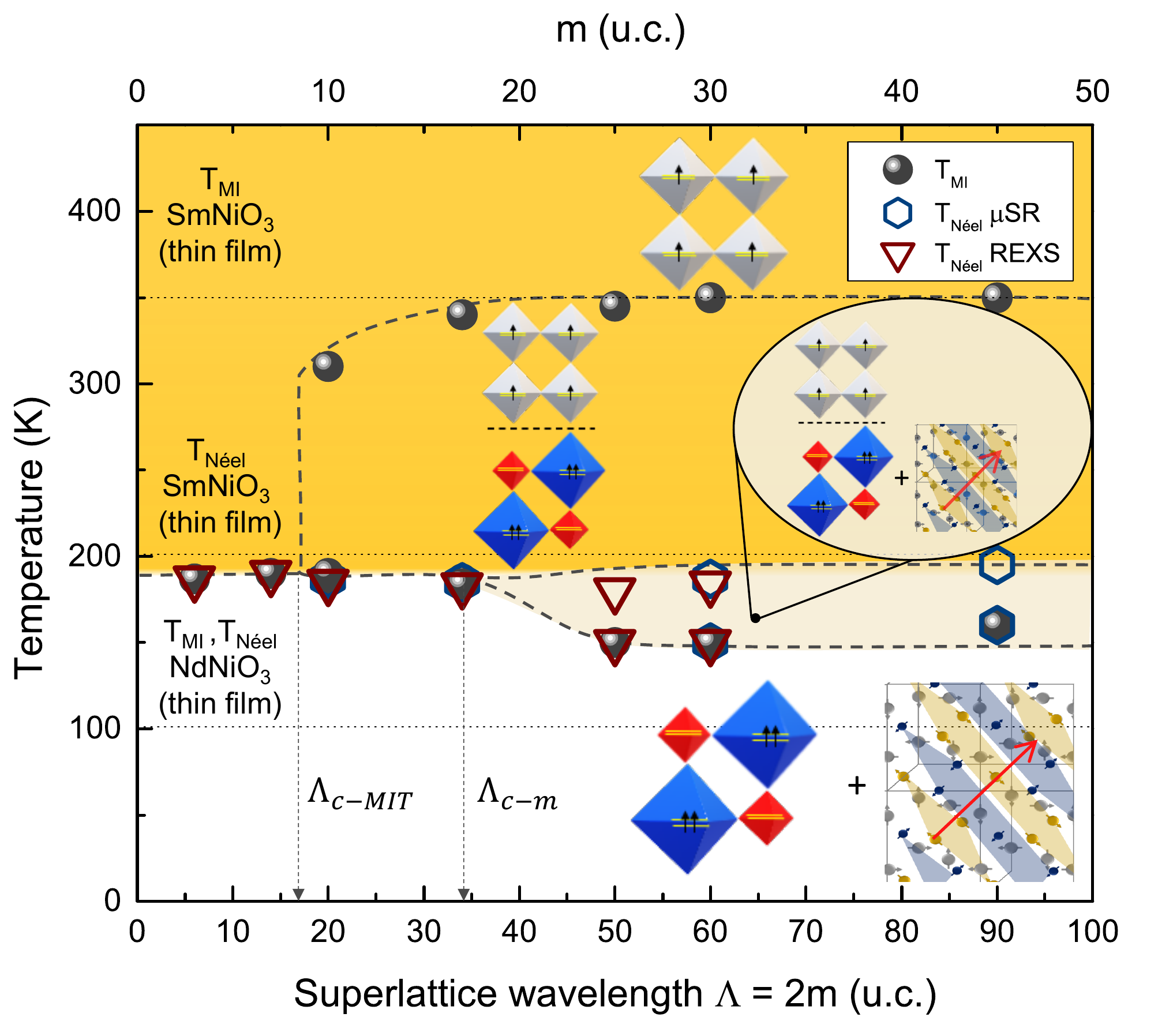}
\caption{Insulator-to-metal (T\textsubscript{MI}) and magnetic (T\textsubscript{Néel}) transition temperatures as a function of the superlattice wavelength $\Lambda$. Yellow indicates the metallic-paramagnetic NdNiO$_3$ and insulating-paramagnetic and metallic-paramagnetic SmNiO$_3$ regions, eggshell corresponds to the region where insulating-antiferromagnetic SmNiO$_3$ and metallic-paramagnetic NdNiO$_3$ layers coexist together, and white refers to the insulating-antiferromagnetic phases present in both compounds at low temperature. The inset sketches show the electronic and structural disproportionation and the antiferromagnetic wave vector in each phase (white, $e_g^1$ proportionate metallic phase; blue, $e_g^2$ B\textsubscript{L}; red, $e_g^0$ B\textsubscript{S} disproportionate insulating phase in the extreme case). B\textsubscript{L} and B\textsubscript{S} are defined as the long bond and short bond Ni sites, respectively. The dashed gray lines serve as guides to the eye. The error bars from the experimental T\textsubscript{MI} and T\textsubscript{Néel} temperatures are smaller than the size of the symbols.}
\label{diagram}
\end{figure}

Figure \ref{diagram} combines together the results obtained from REXS, $\mu$SR and transport measurements as a function of superlattice wavelength $\Lambda = 2m$. The dotted lines indicate the corresponding T\textsubscript{MI} and T\textsubscript{Néel} of 10-nm-thick NdNiO$_3$ and SmNiO$_3$ films grown on (001)\textsubscript{pc}-oriented LaAlO$_3$ substrates. As can be seen in Figure \ref{diagram}, superlattices with lower $\Lambda$, such as (3,3)$_{20}$ and (7,7)$_8$, exhibit electronic properties resembling those of NdNiO$_3$, showing a single insulator-to-metal and magnetic transitions at the same temperature (see Figure S6 (a) in the supplementary material) \cite{RN121}. Superlattices with larger $\Lambda$, i.e. thick NdNiO$_3$ and SmNiO$_3$ layers such as (25,25)$_2$, (30,30)$_2$ and (45,45)$_2$ display two distinct MITs, similar to the insulator-to-metal transitions observed in NdNiO$_3$ and SmNiO$_3$ thin films. Accordingly, two consecutive Néel transitions are seen, i.e. one transition occurring at low temperature, T\textsubscript{Néel} = T\textsubscript{MI} resembling the NdNiO$_3$ layers, and a second one at higher temperature at around 200 K attributed to the SmNiO$_3$ layers, since SmNiO$_3$/LaAlO$_3$ epitaxial films display T\textsubscript{Néel} $<$ T\textsubscript{MI} (see Figure S6 (b) in the supplementary material) \cite{RN217}.  Additionally, there is a region in our phase diagram,  17 u.c. $< \Lambda <$ 34 u.c, with two metal-insulator transitions, although a single magnetic transition is seen to occur together with the lowest T\textsubscript{MI} observed. This behavior has been observed for a (10,10)$_5$ superlattice. A slight kink may be visible in the magnetic volume fraction as a function of temperature for the (17,17)$_3$ superlattice suggesting that this superlattice period may be on the point of the bifurcation in the phase diagram where insulating-antiferromagnetic SmNiO$_3$ layers and metallic-paramagnetic NdNiO$_3$ layers coexist. 

Similar to what is observed in the resistivity measurements, these superlattices display either two magnetic transitions or one depending on the superlattice wavelength. The critical length scale of the antiferromagnetic-paramagnetic phase boundary is around 34 u.c. (13 nm), with a single magnetic transition observed below this magnetic critical wavelength ($\Lambda_{c-m}$). This value is larger than the critical length scale of the metallic-insulating phase separation ($\Lambda_{c-MIT}$ = 16 u.c., 8 nm).

In our previous work, we showed that the length scales of the MIT in SmNiO$_3$/NdNiO$_3$ superlattices can be modeled using a simple Landau theory, including a linear energy difference between the metallic and insulating state for each compound, and an interfacial cost. We refer to the electronic order parameter, the electronic disproportionation, characterizing whether the material is metallic or insulating as \textit{N} \cite{RN290}. Note that, the electronic and structural order parameters are strongly coupled in the nickelates \cite{RN282}, thus, one can integrate out the structural parameter and simply use \textit{N} instead \cite{RNLandscape}. The mismatch cost between the metallic and insulating state is modeled via a gradient squared term. The critical length scale of the antiferromagnetic-paramagnetic phase boundary can also be described by extending the model used previously- where we consider the bulk magnetic energy, a gradient cost between magnetic and non magnetic phases and a coupling term between the electronic and magnetic order parameters. In the insulating state, when NdNiO$_3$ and SmNiO$_3$ layers are both magnetically ordered there will be no phase boundary cost, but when the NdNiO$_3$ layers are paramagnetic and the SmNiO$_3$ layers are antiferromagnetic, a gradient cost will appear. The bifurcation point, $\Lambda_{c-m} = 34$ u.c., is the thickness at which the cost of having an antiferromagnetic-paramagnetic phase boundary is compensated by the bulk energetics. To add magnetism to the bulk Landau theory, we can write:

\begin{multline}
    F(N,m,T)=F_0(N,T)+\frac{1}{2}m^2(B(T)-\alpha N) \\
    +\frac{A}{4}m^4+\frac{\chi_m}{2}(\nabla m)^2 
\end{multline}

\noindent %%To remove the indentation (TAB)
where \textit{N} and \textit{m} are the electronic and magnetic order parameters respectively. The first term depends only on the electronic order parameter \textit{N}, which in SmNiO$_3$ is solely responsible for the MIT, as $m = 0$ throughout the transition. The coupling term $\alpha m^2 N$ couples the magnetic and electronic order parameters, while the rest form a standard second order theory in \textit{m}. Notice the magnetic phase boundary cost $\sim (\nabla m)^2$. When $B(T)-\alpha N < 0$, \textit{m} is non-zero, while for $B(T)-\alpha N > 0$, $m = 0$. Due to the coupling term $\alpha m^2 N $, the magnetic order parameter can be set to $0$ either by changing \textit{N}, or by changing $B(T)$. The parameters for each bulk material then have to be chosen such that in SmNiO$_3$, the Néel transition is driven by a change in $B(T)$ alone, as the two transitions happen at different temperatures, while in NdNiO$_3$ the magnetic order parameter is suppressed by the change in \textit{N}, as the transitions coincide.

To model $F_0(N,T)$ in the simplest possible form that still matches that used in our previous work \cite{RN290}, we pick the following form for the metallic phase:

\begin{equation}
F_0(N,T)=\frac{\chi_M^{-1} N^2}{2}    
\end{equation}

\noindent 
and the insulating phase (above a saturation temperature for SmNiO$_3$):

\begin{equation}
F_0(N,T)=\frac{\chi_I^{-1} (N-N_I)^2}{2}+\Delta(T)    
\end{equation}

\noindent
$\Delta(T)$ depends linearly on temperature, and is equal to $0$ at approximately T = $T_{MI}$. For NdNiO$_3$, $\Delta(T)$ has to be shifted slightly to take into account that the magnetic order parameter raises $T_{MI}$. For SmNiO$_3$, as done in our previous work \cite{RN290} below a saturation temperature, $T_{sat} = 300$ K, the free energy of the insulating state remains constant, i.e. for $T < T_{sat}$, $\Delta(T) = \Delta(T_{sat})$.

\begin{figure}[h!]
\includegraphics[width=\columnwidth]{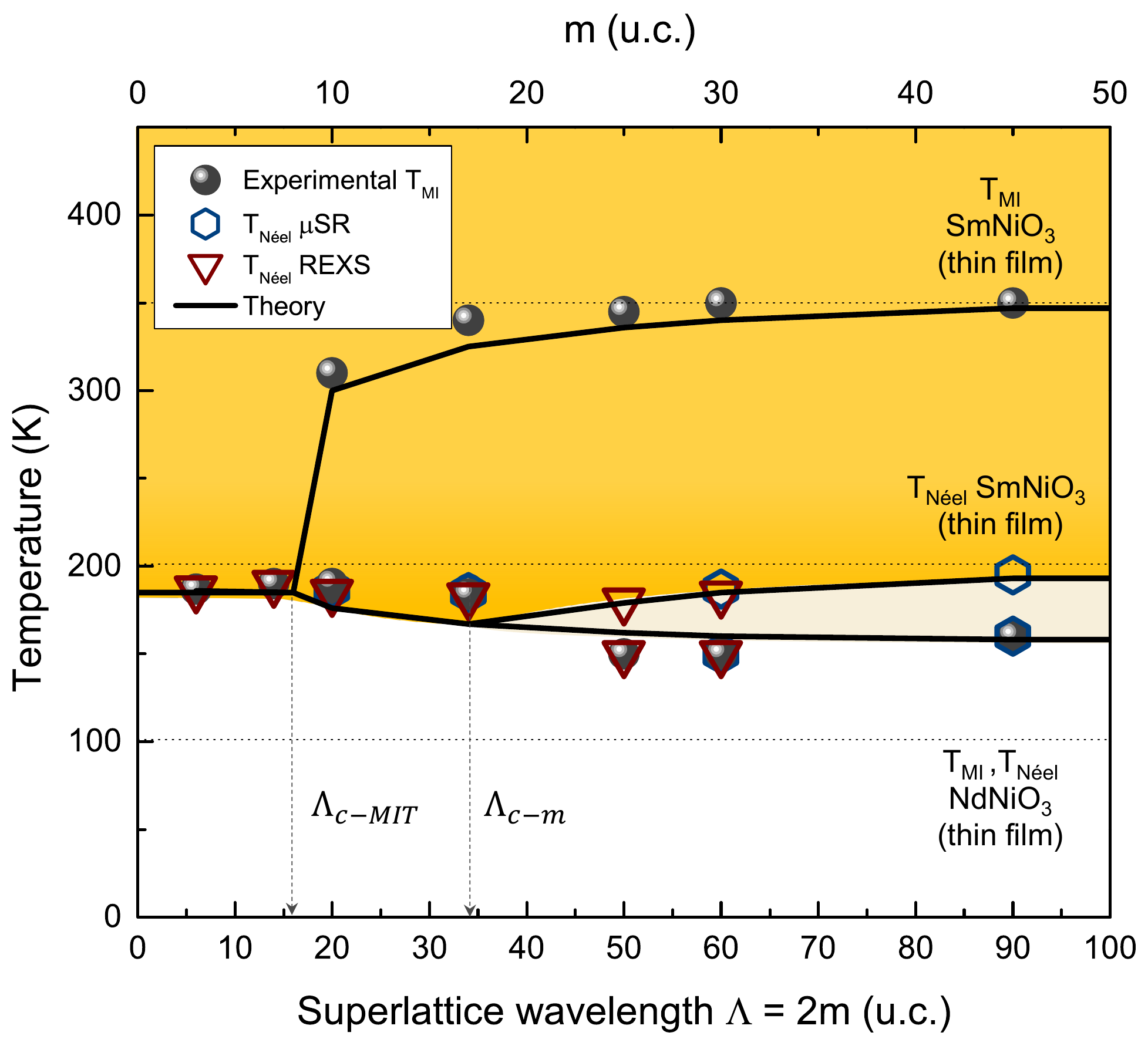}
\caption{Experimental T\textsubscript{MI} and Néel transitions obtained from $\mu$SR and REXS measurements, and output of the Landau model as a function of the superlattice wavelength. The error bars from the experimental transition temperatures are smaller than the size of the symbols.}
\label{magdiagram}
\end{figure}

In order to fully reproduce the transition temperatures for the MIT two points can be used: the critical $\Lambda$ of the MIT, and a single point along the upper branch to determine the temperature at which the free energy difference between the SmNiO$_3$'s insulating and metallic phases reaches a plateau. Importantly, within this full theory, we find that our theoretical landscape also reproduces the length-scale of the MIT, as measured by electron energy-loss spectroscopy in \cite{RN340}. After including the magnetic critical $\Lambda$, we fully reproduce all the higher Néel transitions in the large wavelength regime, as can be seen in Figure \ref{magdiagram}. We find that the quality of the fit is not strongly influenced by the parameters of the fit, provided that it reproduces the magnetic bifurcation point, pointing to a simpler possible relationship between the magnetic energy cost of the mismatch, and the bulk cost.

\begin{figure}[h!]
\includegraphics[width=\columnwidth]{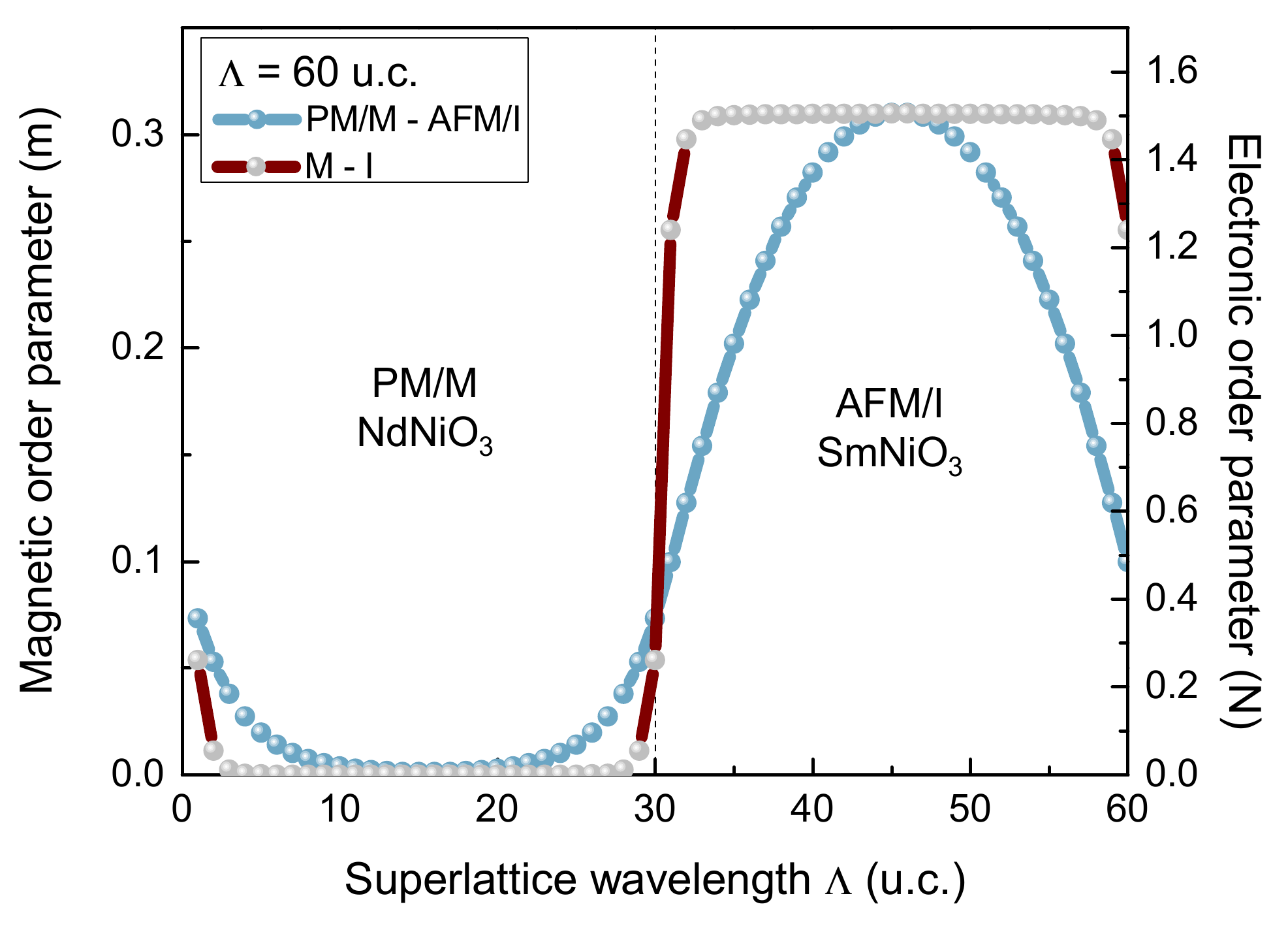}
\caption{Evolution of the electronic order parameter (\textit{N}) across a metallic (M) NdNiO$_3$ and insulating (I) SmNiO$_3$ interface and the magnetic order parameter (\textit{m}) across the paramagnetic-metallic (PM/M) NdNiO$_3$ and antiferromagnetic-insulating (AFM/I) SmNiO$_3$ interface for $\Lambda = 60$ u.c.. Both plots were obtained at 170 K.}
\label{evoordpa}
\end{figure}

Moreover, given the second order nature of the magnetic transition, the evolution of the magnetic order parameter across the paramagnetic/metallic NdNiO$_3$ and the antiferromagnetic/insulating SmNiO$_3$ interface does not show a sharp transition, as compared to the evolution of the electronic order parameter in a first order metal-to-insulator transition, see Figure \ref{evoordpa} (a). 

An examination of the theoretically predicted magnetic profile of a (30,30)$_2$ superlattice ($\Lambda = 60$ u.c.) (Figure \ref{evoordpa}) shows that the magnetic order parameter does not immediately go to zero in the NdNiO$_3$ layers. This behavior is in agreement with the results of the coherence length shown in the REXS section, indicating that the magnetic order ``propagates'' into the NdNiO$_3$ above the Néel transition temperature of the NdNiO$_3$. In other words, the observed behavior might be due to a longer, and more gradual propagation of the magnetic order parameter from one material into the next. %\abg{as dictated by the second order nature of the magnetic transition}.\\

\section{Summary}

Combining a probe of long-range magnetic order such as REXS and a highly sensitive probe of local magnetism like $\mu$SR, we have determined a complete picture of the phase diagram of the SmNiO$_3$/NdNiO$_3$ superlattices. We have confirmed the existence of the antiferromagnetic phase in our superlattices, as observed in the bulk RNiO$_3$ family. The critical length scale over which an antiferromagnetic-paramagnetic phase coexistence occurs is found to be larger than the critical length scale for insulating-metallic phase coexistence. The length scale of the coupling between the antiferromagnetic and paramagnetic phases can be explained in terms of a Landau Theory- where the bulk magnetic energy, a gradient cost between magnetic and non magnetic phases and a coupling term between the electronic and magnetic order parameters are considered. Moreover, the analysis of the coherence length of the magnetic scattering array suggests that when the metallic NdNiO$_3$ layers are assumed to be paramagnetic, more than half of the superlattice still hosts some magnetic order i.e. the magnetic order ``leaks'' into the NdNiO$_3$ layers. Note that in $\mu$SR experiments, the magnetic volumen fraction is above 50$\%$ in the intermediate regime, consistent with a picture where magnetic order is induced, by proximity, in the paramagnetic NdNiO$_3$. For a future study, it may be interesting to perform REXS measurements in superlattices with a different configuration in an effort to understand the potential issue of specific superlattice stacking sequence. This study helps us  understand how distinct magnetic phases couple at interfaces and may be useful for the study of other oxide heterostructures and devices.

\begin{acknowledgments}
We gratefully acknowledge Marco Lopes for technical support and Marios Hadjimichael for useful discussions. This work was supported by the Swiss National Science Foundation through Division II grants nos. 200020 179155, 200020 207338 (C.D., B.M. and J.-M.T) and PP00P2 170564 (M.G.), and has received funding from the European Research Council under the European Union Seventh Framework Programme (FP7/2007–2013)/ERC Grant Agreement no. 319286 (Q-MAC) (C.D., B.M. and J.-M.T). 

\end{acknowledgments}

\nocite{*}

\bibliography{apssamp}% Produces the bibliography via BibTeX.

\end{document}